\documentclass[sigconf]{acmart}

\AtBeginDocument{}

\copyrightyear{2026}
\acmYear{2026}
\setcopyright{cc}
\setcctype{by}
\acmConference[MIG '26]{The 19th ACM SIGGRAPH Conference on Motion, Interaction, and Games}{December 11--13, 2026}{Charleston, SC, USA}
\acmBooktitle{The 19th ACM SIGGRAPH Conference on Motion, Interaction, and Games (MIG '26), December 11--13, 2026, Charleston, SC, USA}
\acmDOI{10.1145/3828647.3857300}
\acmISBN{979-8-4007-2824-2/2026/12}

\begin{document}

\title{CrowdDirector: Runtime Orchestration of Crowd Simulation through Natural-Language Directives}

\author{Xiaohan Sun}
\affiliation{%
  \institution{Trinity College Dublin}
  \city{Dublin}
  \country{Ireland}}
\email{sunx4@tcd.ie}

\author{Vuk Radovanovic}
\affiliation{%
  \institution{Trinity College Dublin}
  \city{Dublin}
  \country{Ireland}}
\email{radovanv@tcd.ie}

\author{Merce Alvarez}
\affiliation{%
  \institution{CYENS - Centre of Excellence}
  \city{Nicosia}
  \country{Cyprus}}
\email{m.alvarez@cyens.org.cy}

\author{Carol O'Sullivan}
\affiliation{%
  \institution{Trinity College Dublin}
  \city{Dublin}
  \country{Ireland}}
\email{carol.osullivan@tcd.ie}

\author{Yiorgos Chrysanthou}
\affiliation{%
  \institution{CYENS - Centre of Excellence}
  \city{Nicosia}
  \country{Cyprus}}
\email{y.chrysanthou@cyens.org.cy}

\author{Panayiotis Charalambous}
\affiliation{%
  \institution{CYENS - Centre of Excellence}
  \city{Nicosia}
  \country{Cyprus}}
\email{p.charalambous@cyens.org.cy}

\renewcommand{\shortauthors}{Sun et al.}

\begin{abstract}
Natural language offers an intuitive way to direct a virtual crowd while a
simulation is running. Handwritten rules execute efficiently but cannot respond
flexibly to varied instructions, whereas asking a large language model (LLM) to
control the crowd at every update introduces repeated latency, cost, and the
risk of unsupported actions. We present CrowdDirector, a framework that uses an
LLM only when the user adds or changes an instruction. The LLM translates the
request into a structured instruction, and a scene validator checks whether
the entities and actions it references exist in the current Unity scene. Accepted instructions are stored and remain continuously active as the simulation is running. At each simulation tick, a graph neural network uses the current scene state and active instruction to choose an available action and target for each agent, while Unity handles movement, interactions, and animation. The network is trained offline on simulated scenarios, with LLM guidance on how agents should respond to instructions. Our evaluations show the scene validator's feasibility
decisions matched human judgements more often than the LLM's own assessment,
and on a behavioural benchmark CrowdDirector follows user instructions more closely than handwritten
rules while avoiding repeated LLM calls. It therefore provides a practical
middle ground between fixed rules and continuous LLM control.
\end{abstract}

\ccsdesc[500]{Computing methodologies~Multi-agent systems}
\ccsdesc[500]{Computing methodologies~Intelligent agents}
\ccsdesc[300]{Computing methodologies~Animation}
\ccsdesc[300]{Human-centered computing~Interactive systems and tools}

\keywords{crowd simulation, large language models, graph neural networks, runtime authoring, intelligent virtual agents, interactive storytelling}

\begin{teaserfigure}
  \centering
  \includegraphics[width=\textwidth]{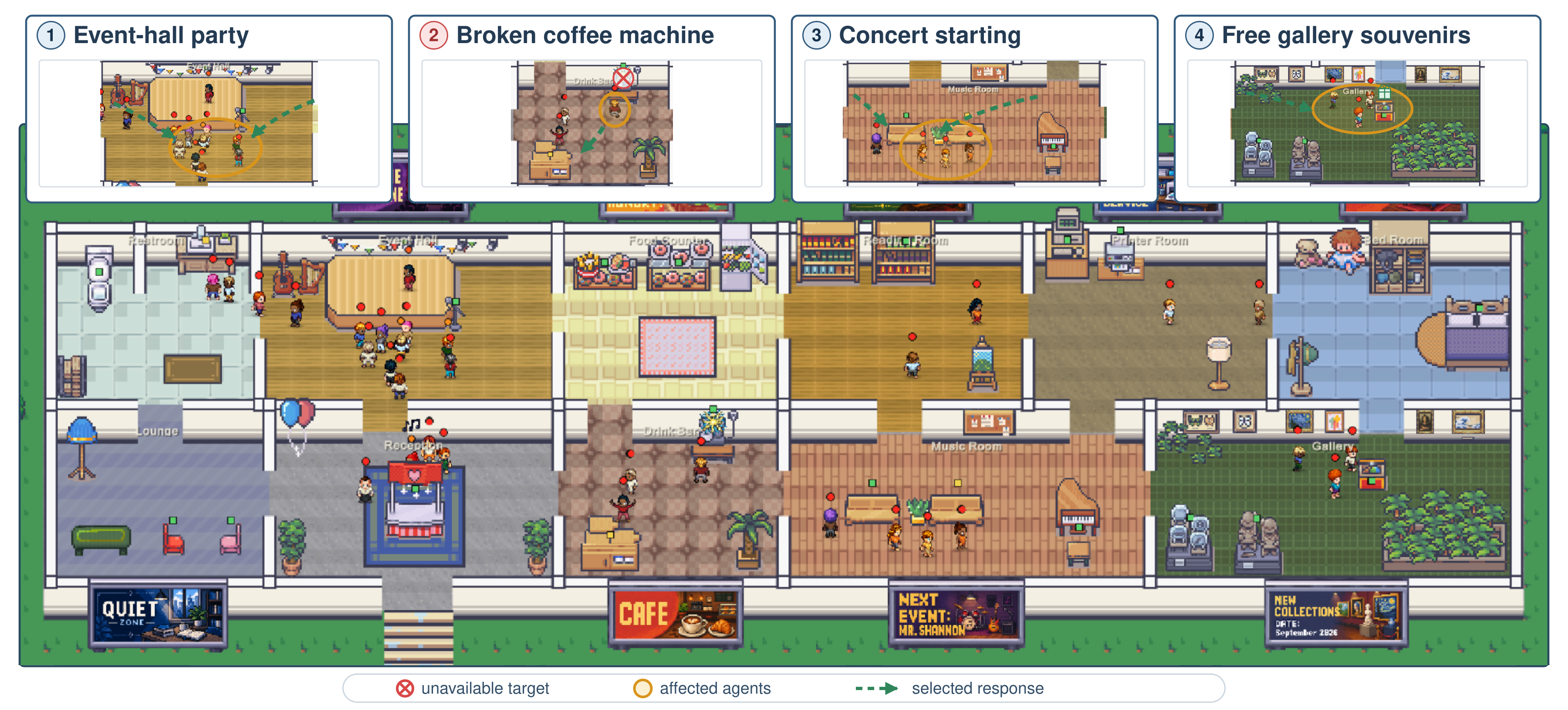}
  \caption{CrowdDirector directs a running crowd with natural language. Four
instructions are active at once: visitors gather for a party in the Event Hall,
agents waiting at the broken coffee machine ($\otimes$) are redirected to
another drink source, visitors collect in the Music Room before a concert, and
interested visitors move to the Gallery. Circles mark the agents each
instruction affects. Each instruction is interpreted once, then applied locally
without further language inference.}
  \Description{A top-down view of a 2D venue in which four natural-language
  instructions are active at once, with coloured circles marking the agents each
  one affects. Four inset panels above show close-ups of the event hall, the
  broken coffee machine, the music room, and the gallery.}\vspace{2em}
  \label{fig:teaser}
\end{teaserfigure}

\maketitle

\section{Introduction}
\label{sec:intro}

Crowd simulation is commonly organized as a hierarchy. Global planning selects
routes, local navigation avoids collisions, and behavioural logic determines
which goals and interactions agents should pursue~\cite{reynolds1987flocks,
helbing1995social,vandenberg2011orca,treuille2006continuum}. This separation
supports real-time animation, but the behavioural layer remains expensive to
author. For every event, a designer must decide in advance who is affected,
where those characters should go, and how their current activity is interrupted
or resumed. These requirements become restrictive in interactive
scenes, where the desired response may be introduced only after the simulation
has started.

Natural language is an attractive runtime interface because a single
instruction can express intent, scope, and context: for example, ``staff gather
at the entrance,'' ``the coffee machine is broken; redirect anyone waiting for
a drink,'' or ``keep visitors away from the spill.'' Recent systems use
language models to generate agent plans, crowd scenarios, motion, dialogue, and
animation~\cite{park2023generative,ji2024textcrowd,cao2026crowdmogen,
li2024words,liu2025emergent}. Direct LLM control, however, is difficult
to place inside a repeated crowd-control loop. The model may reference
objects or actions that do not exist in the scene, or select the wrong agents.
Calling it at every update is also slow, costly, and nondeterministic, and its
output must still respect simulator constraints such as capacity and
navigation.

This leaves a practical gap. A runtime authoring system must understand a new
instruction, determine whether it can be carried out in the current scene, keep
it active for the requested duration, and turn it into decisions for many
agents. Handwritten rules make these decisions efficiently, but only
for responses that were authored in advance. Per-tick LLM control can handle
varied instructions, but querying the model at every update multiplies latency
and API cost by the length of the session, even when the user's
intent has not changed.

Our central premise is that the LLM should interpret a new instruction, not
control every agent at every update. CrowdDirector therefore calls the LLM
only when the user introduces or revises an instruction, and only to convert the
text into a structured form. Each new instruction then passes through
three steps: (1)~a fixed, programmatic \emph{scene validator} resolves the named
agents, zones, and objects, checks whether the requested actions exist, and
reports whether all, part, or none of the instruction can be executed; (2)~an
accepted instruction is stored as a \emph{directive}, a checked instruction that
remains part of the simulation state while it is active; and (3)~at each later update, a graph neural network policy (graph policy) chooses an available action and target for each agent.

The graph policy is trained offline on generated
scenarios. For every decision, short simulation rollouts estimate how well each
available action would turn out, which gives a base training label. When a
decision is related to an instruction, an LLM additionally ranks the same list
of available actions, and this ranking is blended with the base label using
fixed weights (Section~\ref{sec:supervision}). This design uses language to guide responses to an
instruction without allowing the LLM to add actions or targets that the scene
does not support.

Our contributions are:
\begin{enumerate}
  \item A runtime crowd-control framework integrated with Unity that interprets
  a natural-language instruction once, checks it against the current scene,
  and stores it as an active part of the simulation instead of calling an LLM
  at every update.

  \item A graph policy that chooses from the actions and targets available to
  each agent. It is trained from simulation rollouts, supplemented by
  constrained LLM rankings for instruction-related responses.

  \item An evaluation against handwritten rules and per-tick LLM
  control. It measures whether instructions are checked correctly, whether the right agents respond, and how much each update costs in latency and API calls.
\end{enumerate}

\vspace{1em}
\section{Related Work}
\label{sec:related}

\subsection{Crowd Simulation and Authoring}
Foundational crowd models generate collective motion through local interaction
rules, continuous forces, reciprocal avoidance, and aggregate flow
fields~\cite{reynolds1987flocks,helbing1995social,vandenberg2008rvo,
vandenberg2011orca,treuille2006continuum,narain2009aggregate}. Empirical work
has further identified compact interaction laws for pedestrian
motion~\cite{karamouzas2014powerlaw}. These approaches primarily determine
\emph{how} agents move once a goal is known. Behavioural simulation adds
perception, roles, schedules, and cognitive state to determine \emph{what}
agents do~\cite{funge1999cognitive,musse2001hierarchical,sung2004scalable,
shao2005autonomous,pelechano2007hidac}.

Crowd authoring systems expose higher-level controls through spatial painting,
behaviour trees, semantic environments, and smart objects~\cite{ulicny2004crowdbrush,
shoulson2011pbt,kallmann1999objects}. Recent learned approaches provide
controllable behaviour profiles, example-driven policies, or interpretable
mesoscopic parameters~\cite{panayiotou2022ccp,panayiotou2025cedrl,
lemonari2025mpact}. LexiCrowd adds natural-language authoring by learning a
mapping from textual descriptions to continuous simulator parameters for goal
seeking, grouping, and interaction with areas of interest; its training data
consist of LLM-generated text--behaviour pseudo-pairs~\cite{lemonari2024lexicrowd}.
Its output is therefore a crowd-level behaviour profile. Our focus is different. We address live edits: an instruction arrives while the simulation is running and refers to the entities present in the scene at that time.

\subsection{Language-Based Control of Virtual Agents and Crowds}
LLMs have enabled agents that plan, remember, converse, and acquire skills.
Generative Agents store memories and use them to plan long-horizon
social behaviour~\cite{park2023generative}; Voyager uses repeated LLM calls to
acquire executable skills~\cite{wang2024voyager}; and SayCan combines language
with a separate estimate of executable affordances~\cite{ichter2023saycan}.
Shan and Michel similarly combine an LLM with goal-oriented action planning:
natural-language player commands change the priorities of predefined NPC goals,
and the planner constructs an executable response~\cite{shan2024goap}. These works use a language model to reason on behalf of a single agent.

In crowd animation, Text-Guided Synthesis of Crowd Animation generates group
distributions and velocity fields~\cite{ji2024textcrowd}; CrowdMoGen uses an
LLM-assisted planner for collective motion generation~\cite{cao2026crowdmogen};
and Gen-C uses LLM-generated scenarios to train graph generative models of
crowd actions and interactions~\cite{panayiotou2026genc}. These methods map
language to a generated animation, motion pattern, or reusable crowd model.

CrowdVLA instead places language-conditioned reasoning inside the runtime
controller. It models each pedestrian as a vision--language--action agent that
uses an agent-centric visual observation to select one of 64 short-horizon
motion skills and replans after executing each skill~\cite{hwang2026crowdvla}.
Its goal is context-aware navigation and social-norm compliance through an
end-to-end visual-language policy. Other runtime systems query agent-centric
LLMs to couple dialogue, social state, and navigation~\cite{liu2025emergent,
wang2025peba}. CrowdDirector instead interprets a crowd-level instruction once when it arrives, and the resulting directive conditions later graph-policy decisions without further language inference.

\subsection{Graph-Based Policies and LLM Training Labels}
Interactive scenes are naturally relational: agents belong to groups, occupy
zones, reserve objects, and respond to operations targeting other entities.
Graph networks provide an inductive bias for this structure~\cite{battaglia2018relational},
while relational graph convolutions distinguish typed edges~\cite{schlichtkrull2018rgcn}.
Pointer-style prediction is suitable when outputs select from a variable set of
available choices~\cite{vinyals2015pointer}. We combine these ideas by
propagating state over a crowd graph and scoring the candidate actions available
to each agent rather than predicting from a fixed action vocabulary alone.

Imitation learning and policy distillation train a policy from the
decisions of a more expensive
teacher~\cite{ross2011dagger,rusu2015policy,hinton2015distilling}. CrowdVLA
applies a related idea to crowd simulation, using simulator rollouts to
supervise the counterfactual outcomes of alternative motion
skills~\cite{hwang2026crowdvla}. We follow the same approach, using simulation rollouts together with a language model ranking of the available actions as training labels for our policy.

\section{Method}
\label{sec:method}

\subsection{System Overview}
\label{sec:method-overview}

\begin{figure*}[t]
  \centering
  \includegraphics[width=\textwidth]{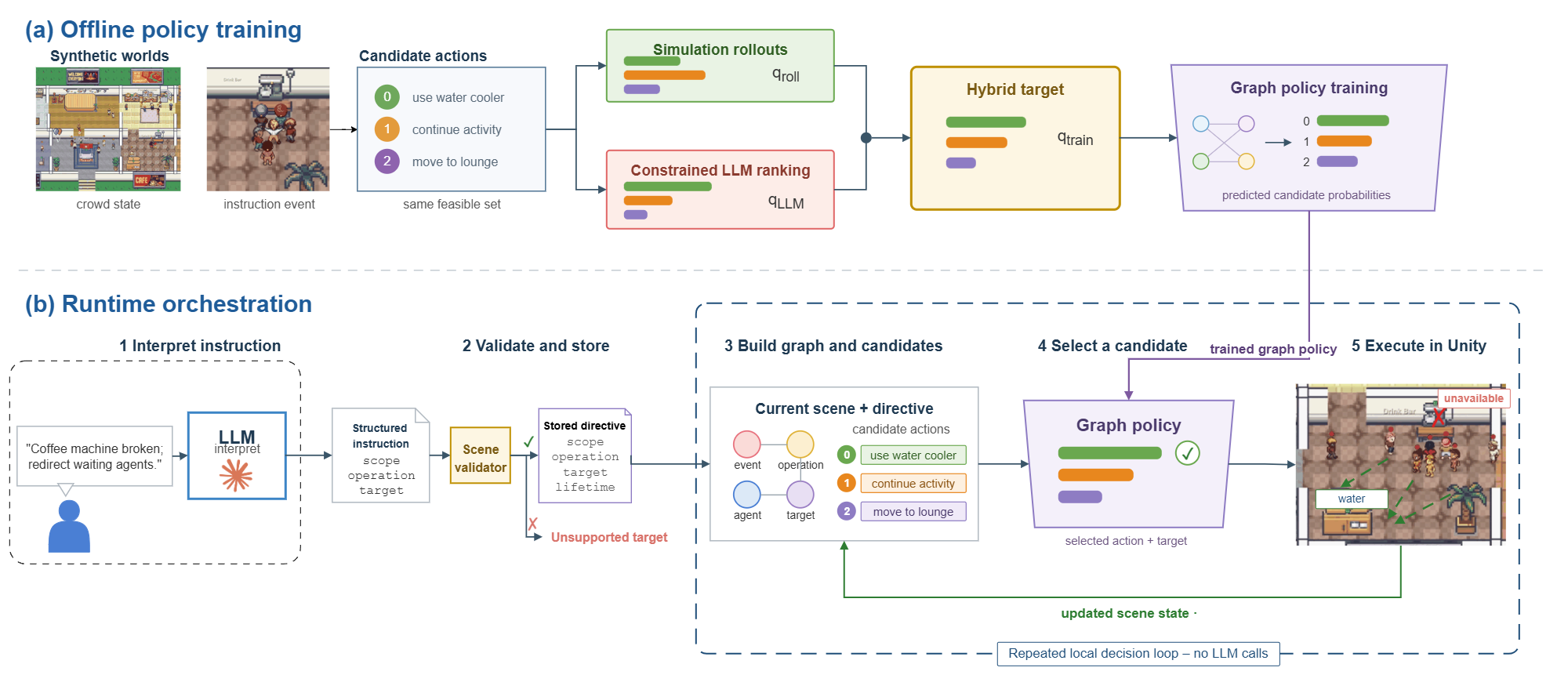}
  \caption{Overview of CrowdDirector. \textbf{(a)~Offline policy
  training.} In generated worlds, simulation rollouts and a constrained LLM
  ranking score the same set of available actions, giving
  $q_{\mathrm{roll}}$ and $q_{\mathrm{LLM}}$. Averaging them yields the hybrid
  target $q_{\mathrm{train}}$, which trains the graph policy. Because both
  distributions cover the same candidates, the blend cannot introduce an
  unavailable action. \textbf{(b)~Runtime orchestration.} An LLM interprets a
  new instruction once~(1), and the scene validator checks it against the live
  scene, storing an accepted result as a directive and rejecting an unsupported
  target~(2). The current scene and active directives are then rebuilt as a
  crowd graph with a candidate list for each agent~(3), the trained policy
  selects one candidate per agent~(4), and Unity executes the result~(5).}
  \Description{A two-part block diagram. The upper part shows offline training,
  where simulation rollouts and a constrained LLM ranking of the same candidate
  actions are averaged into the target that trains the graph policy; the lower
  part shows the five runtime stages, from interpreting an instruction through
  scene validation, graph construction, and policy selection to execution in
  Unity.}
  \label{fig:pipeline}
\end{figure*}

CrowdDirector applies an instruction to a running crowd simulation in five
stages (Figure~\ref{fig:pipeline}):
\begin{enumerate}
  \item an LLM converts the raw instruction into a structured instruction;
  \item the scene validator checks that instruction and stores an accepted
  result as a directive;
  \item the current scene and directive are represented as a crowd graph, and
  a candidate generator lists the actions available to each agent;
  \item the graph policy selects one candidate action for each agent; and
  \item the Unity execution layer carries out the selected actions and requests
  replanning when necessary.
\end{enumerate}

Only the first stage requires language inference, and it runs only when an
instruction is introduced or revised. The remaining stages form the local
runtime loop. We use one running example throughout this section: in a cafe,
the user states, ``The coffee machine is broken; redirect anyone waiting for a
drink.''

\subsection{Translating and Checking an Instruction}
\label{sec:grounding}

The LLM (Claude Haiku) \cite{anthropic2024claude3} converts a user instruction into a compact structured instruction. It proposes one or more \emph{operations}, each a requested change
to scene state, such as disabling the drink affordance of the coffee machine. For the running example, a shortened JSON record is:

\begin{quote}\small\ttfamily\setlength{\parindent}{0pt}
\{\\
"operation": "disable\_affordance",\\
"target": \{"object": "coffee\_machine\_1",\\
"affordance": "drink"\}\\
\}
\end{quote}
This example and the two later examples omit internal identifiers and timestamps
for readability. The
structured instruction is a claim about the user's intent, not permission to change the
scene.

The scene validator resolves every referenced entity against the live
simulation. For each name, it finds the matching entity in the current
scene and checks that this entity supports the requested action. It returns one of five states:

\begin{itemize}
  \item \emph{executable}: the complete request is valid;
  \item \emph{partially executable}: a supported subset can be applied;
  \item \emph{unsupported}: the requested behaviour is outside the simulator
  vocabulary;
  \item \emph{ambiguous}: several plausible groundings remain; and
  \item \emph{invalid reference}: a named entity does not exist or has the
  wrong type.
\end{itemize}

In the running example, the
validator resolves the coffee machine, verifies its \emph{drink} affordance, and
accepts the operation. Its shortened output makes the decision explicit:

\begin{quote}\small\ttfamily\setlength{\parindent}{0pt}
\{\\
"status": "executable",\\
"resolved\_target": "coffee\_machine\_1:drink",\\
"scope": "all"\\
\}
\end{quote}

Only executable structured instructions, or partial instructions explicitly accepted by the
user, modify the simulation. This boundary prevents the LLM from inventing an
unsupported action or silently bypassing scene constraints. It also separates
two questions that an LLM can otherwise conflate: what the user appears to
mean, and what the current simulator can actually execute.

After validation, the resolved instruction is stored as a directive. It records
who is affected, what scene change is requested, and when the
request should stop. The directive is not a copy of the structured
instruction: it restates the request as a set of scene patches, and one
requested operation may expand into several. An abridged form is:
\begin{quote}\small\ttfamily\setlength{\parindent}{0pt}
\{\\
"intent": "object\_disable",\\
"scope": "all",\\
"operations": ["disable drink at coffee\_machine\_1",\\
"enable drink in lounge",\\
"raise lounge attraction"],\\
"lifetime": "60 ticks"\\
\}
\end{quote}
The directive does not name the agents it affects. It changes the scene rather
than selecting a group, and which agents respond is left to the policy.
The candidate generator and graph policy read all active
directives at every decision update. The coffee-machine directive therefore
continues to redirect agents who become thirsty later, without another LLM
call. A higher-priority interruption may temporarily override it; when that
interruption ends, the original directive becomes active again.

\subsection{Representing the Current Crowd Scene}
\label{sec:representation}

The current scene is encoded as a crowd graph with six node
types:
\begin{itemize}
  \item \emph{agents}, \emph{zones}, \emph{objects}, and \emph{groups}, which
  describe the scene itself; and
  \item \emph{events} and \emph{operations}, which describe active directives.
  An event node records an active situation, and
  its operation nodes record the requested scene changes.
\end{itemize}
Directed relations connect these nodes. They record which zone an agent is in, which group it belongs to, which affordances each object offers, and which agents an operation applies to. The graph is rebuilt at every decision update, so a new instruction adds only its own event and operation nodes, together with the relations linking them to the agents and objects it names.

A fixed, Maslow-inspired rule~\cite{maslow1943theory} sorts an agent's nine needs into three
tiers, from basic physical needs up to social and self-fulfilment needs, so an
urgent lower-tier need is addressed before a higher-tier one. Needs grow over
time and fall when the agent performs a relevant action. They follow fixed
simulation rules and are not learned, but they form part of the agent's node in
the graph.

For each agent, the candidate generator then lists at most 64
actions the agent could take now, drawn from the scene, these needs, and the
active directives. Table~\ref{tab:actions} gives the action types. Each listed
action also records its target and how easy that target is to reach and use,
such as its distance and whether it still has room.

\begin{table}[t]
  \centering
  \caption{Action types the candidate generator can list, with examples
  from the running example. The actions available to an agent depend on what
  the current scene provides.}
  \label{tab:actions}
  \small
  \setlength{\tabcolsep}{4pt}
  \begin{tabular}{lll}
    \toprule
    Action & Target & Example \\
    \midrule
    walk    &  zone           & walk to the lounge \\
    leave   &  exit          & leave by the fire exit \\
    drink   &  object        & drink at the water cooler \\
    rest    &  object        & rest on the sofa \\
    work    &  object        & work at a desk \\
    observe &  object        & look at the whiteboard \\
    idle    &  agent  & wait in place \\
    \bottomrule
  \end{tabular}
\end{table}

Hard structural impossibilities are removed before inference. For example, a
candidate is excluded when its action requires an affordance that the target type
cannot provide. Conditions that may change at runtime, such as a
disabled machine or a temporary hazard, are kept in the candidate features
instead of being filtered out. The policy can therefore learn to avoid such
targets and to reconsider them once the condition clears.

Some directives assign a specific destination, as in
``staff gather at the entrance.'' The entrance zone already appears in the graph
and in each staff member's candidate list, but nothing yet tells the policy
which candidate the directive is pointing at. We therefore mark it with a
single binary \emph{directive-target} feature: it is 1 for the candidate that
matches the agent's assigned destination and 0 for every other candidate. This
is a marker, not a priority: it does not add an action or force its selection,
and the policy learns during training how strongly to follow it.

\subsection{Choosing an Action for Each Agent}
\label{sec:policy}

\begin{figure*}[t]
  \centering
  \includegraphics[width=\textwidth]{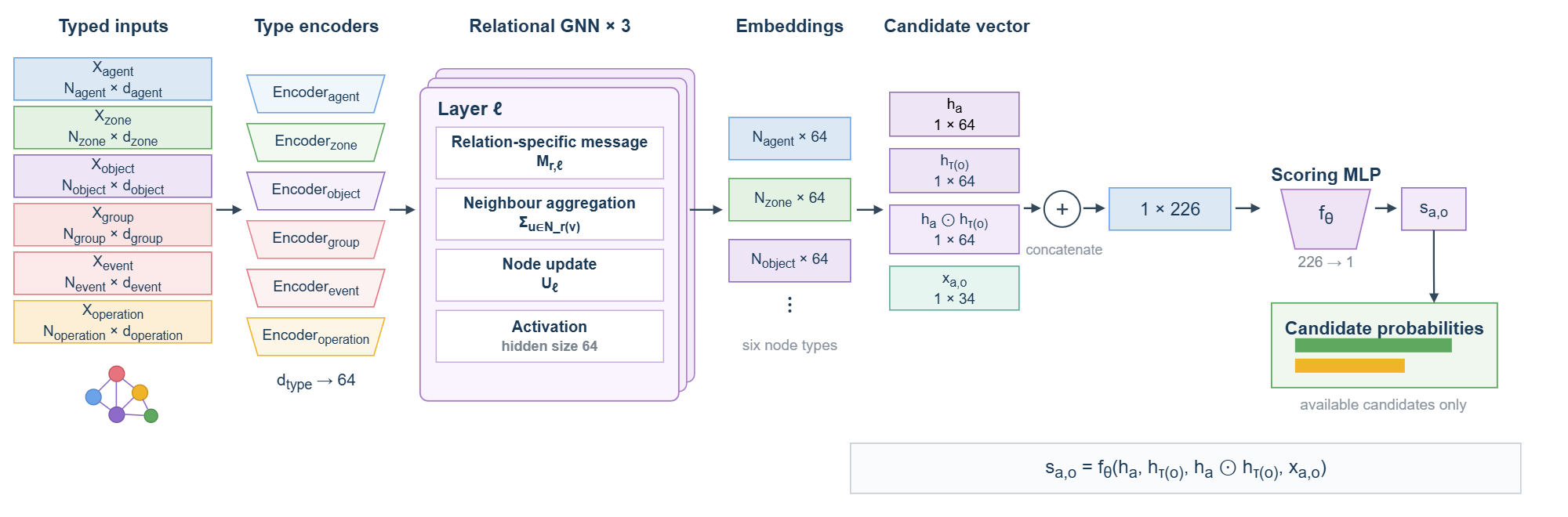}
  \caption{Architecture of the graph policy. Type-specific encoders produce
  64-dimensional node representations, which are updated by three relational
  GNN layers. A shared multilayer perceptron scores each candidate from its
  agent representation, target representation, their interaction, and the
  34-dimensional candidate features.}
  \Description{A left-to-right architecture diagram in which six typed node
  inputs pass through type-specific encoders and three relational graph neural
  network layers. A shared multilayer perceptron then scores each candidate
  action from the agent representation, the target representation, their
  element-wise product, and 34 candidate features.}
  \label{fig:network}
\end{figure*}

The graph policy uses a three-layer relational graph neural network with a
64-dimensional hidden representation (Figure~\ref{fig:network}). Per-type encoders first project the six
node types into the shared space. Relation-specific message passing then
propagates information between connected entities. Each message-passing
layer moves information one edge further through the graph, so the number of
layers equals the maximum path length information can travel. Three layers
allow an agent to receive directive information along three-edge paths such as
\[
\text{event}\rightarrow\text{operation}\rightarrow\text{target}
\rightarrow\text{agent}.
\]
The graph policy contains no text encoder: natural language has already been
converted into resolved operations and explicit graph state before inference.

For each candidate action $o$ of agent $a$, the policy scores
every entry of the agent's candidate list and normalizes those scores into a
distribution over the list itself. This follows the idea of pointer
networks~\cite{vinyals2015pointer}: the output is a selection from the input
list, so the number of choices can differ per agent and per update, unlike a
classifier with a fixed set of output classes. The scorer combines the agent embedding, target embedding, their element-wise interaction, and the 34 candidate features into a single 226-dimensional vector:
\begin{equation}
s_{a,o}=f_{\theta}\!\left(
h_a,\,h_{\tau(o)},\,h_a\odot h_{\tau(o)},\,x_{a,o}
\right),
\label{eq:score}
\end{equation}
where $h_a$ is the encoded agent state, $h_{\tau(o)}$ is the encoded target,
$x_{a,o}$ is the candidate feature vector, and $f_{\theta}$ is a multilayer
perceptron. A masked softmax over the candidate actions produces the final
action distribution.

In the running example, the directive has disabled the broken
coffee machine's drink affordance, so the water cooler is the remaining way for
a thirsty agent to get a drink. The graph policy sees only actions supported by
the scene and selects the water cooler for the affected agent:
\begin{quote}\small\ttfamily\setlength{\parindent}{0pt}
\{\\
"agent": "agent\_07",\\
"candidate\_actions": [\\
~~~~"use coffee\_machine\_1 (unavailable)",\\
~~~~"use water\_cooler\_1",\\
~~~~"rest"],\\
"selected\_action": "use water\_cooler\_1"\\
\}
\end{quote}

\subsection{Creating the Training Labels}
\label{sec:supervision}

This stage is offline: it produces the labels used to train the graph
policy, and does not run during the loop above. The graph policy learns a probability distribution over each agent's candidate
actions. Simulation rollouts generate a base label for every decision. For instruction-related
decisions with a valid LLM ranking, the rollout and language distributions are
averaged; routine decisions and invalid or missing LLM outputs use the rollout
distribution alone.

\paragraph{Routine decisions.}
For every candidate action, the data generator clones the headless
simulation and performs $K=8$ perturbed rollouts over a horizon of $H=5$
committed steps. The return is dominated by need relief and event
progress, and penalizes costs such as travel delay and failed actions. Returns
are standardized within the candidate set and converted into a soft target
distribution $q_{\mathrm{roll}}$. This distribution reflects consequences
that are difficult to infer from a static snapshot, including changing needs,
capacity, travel cost, and contention.

\paragraph{Instruction-related decisions.}
For each world containing an active instruction, one temperature-zero LLM call labels all
agents in that world. The prompt describes the active operations, the
current scene state, each agent's most urgent needs, and each agent's numbered
candidate list. The LLM returns the top three candidate indices for each
agent. We assign probability masses of $0.60$, $0.25$, and $0.15$ to the
first-, second-, and third-ranked candidates, respectively, forming
$q_{\mathrm{LLM}}$. Invalid indices, missing predictions, and unparseable
outputs are discarded; in those cases, the simulation label is retained. The
LLM can therefore indicate which available response best follows the
instruction, but cannot invent a new behaviour. The ranking weights express a
strong first preference while retaining two alternatives; they are a fixed
data-generation choice rather than a tuned model parameter.

The final training target is a hybrid of the two distributions:
\begin{equation}
q_{\mathrm{train}}=
\begin{cases}
(q_{\mathrm{roll}}+q_{\mathrm{LLM}})/2,
  & \text{valid LLM ranking},\\
q_{\mathrm{roll}}, & \text{otherwise}.
\end{cases}
\label{eq:training-target}
\end{equation}
This blend retains information about travel, capacity, and contention from the
rollouts while shifting probability toward actions that better follow the
instruction. Both distributions follow the same candidate order, so their
average cannot introduce an unavailable action.

Table~\ref{tab:training-example} shows an abridged instruction-related
training record. The LLM call contains the candidate lists for all agents in
the world, but only one agent is shown here for clarity. Candidate indices are
created by the simulator before the LLM is queried. The table shows the LLM
component numerically; the final training target also includes the rollout
component as defined in Equation~\ref{eq:training-target}.

\begin{table}[t]
  \centering
  \caption{Abridged example of an instruction-related LLM label.
  Probabilities follow the displayed candidate order.}
  \label{tab:training-example}
  \small
  \setlength{\tabcolsep}{4pt}
  \begin{tabular}{p{0.30\columnwidth}p{0.60\columnwidth}}
    \toprule
    Field & Example \\
    \midrule

    Instruction &
    ``The coffee machine is broken; redirect anyone waiting for a drink.'' \\

    Relevant state &
    \texttt{coffee\_machine\_1: broken};

    \texttt{agent\_07: waiting, thirst high} \\

    Candidate 0 &
    \texttt{use water\_cooler\_1} \\

    Candidate 1 &
    \texttt{continue current\_activity} \\

    Candidate 2 &
    \texttt{move to lounge} \\

    LLM ranking &
    $[0,\,2,\,1]$ \\

    LLM label &
    $q_{\mathrm{LLM}}=[0.60,\,0.15,\,0.25]$ \\

    \bottomrule
  \end{tabular}
\end{table}

\paragraph{Training corpus and optimization.}
We generate the training worlds procedurally, varying the layout, the
number of agents, and the instruction that arrives. They cover ordinary
behaviour, where agents simply satisfy their own needs, and the instruction
types used in evaluation, such as a broken machine, an evacuation, and a
gathering. The corpus holds 2,776 worlds and 73,606 agent decisions in total:
2,148 worlds train the policy, 258 select the final checkpoint, and 370 are
held out to test combinations of instructions the policy never saw during
training. An LLM ranking is available for 61,439 of the decisions
(Table~\ref{tab:dataset}).

\begin{table}[t]
  \centering
  \caption{Generated training corpus. Validation is used for checkpoint
  selection; composition splits are held out.}
  \label{tab:dataset}
  \small
  \setlength{\tabcolsep}{3.0pt}
  \resizebox{\columnwidth}{!}{%
  \begin{tabular}{llrrr}
    \toprule
    Split & Use & Worlds & Decisions & LLM-blended \\
    \midrule
    full/train       & train     & 1,280 & 40,283 & 35,115 \\
    full/validation  & selection &   258 &  8,340 &  7,369 \\
    full/composition & held out  &   268 &  8,406 &  8,306 \\
    ext/train        & train     &   584 &  9,483 &  4,624 \\
    ext/composition  & held out  &   102 &  1,685 &    616 \\
    new-event/train  & train     &   284 &  5,409 &  5,409 \\
    \midrule
    Total            &           & 2,776 & 73,606 & 61,439 \\
    \bottomrule
  \end{tabular}}
\end{table}

Training minimizes the Kullback--Leibler (KL) divergence between the
policy's distribution over an agent's candidates and the training target. KL
divergence measures how far one probability distribution is from another, so
minimizing it makes the policy's choices match the target. Additional ranking terms encourage the policy to prefer actions that relieve a need, respond to an active instruction, choose a working alternative when the usual one is unavailable, or continue an activity the agent has already started. We
use Adam with batch size 64 and an initial learning rate of
$3\times10^{-3}$, and keep the checkpoint with the lowest validation
KL under early stopping.

\subsection{Executing the Selected Actions in Unity}
\label{sec:execution}

The policy decides \emph{what} an agent should attempt; the Unity runtime
determines \emph{how} that decision is executed. Our implementation splits this across two
processes. A Python process holds the symbolic state, meaning agent needs,
active directives, and the crowd graph, and runs the trained policy. A Unity 2D
client holds the scene geometry and executes each selected action. The two exchange messages over a WebSocket: Unity
reports what has happened in the scene, and Python replies with the action each
agent should attempt next.

After a level is instantiated, Unity reports which objects and zones actually
exist, where each object can be used from, and which of them can be reached.
The policy draws its targets from that report, so an object that was
never placed, or that cannot be reached, is never offered as an action. Unity
then handles path planning and collision avoidance, object interactions, and
animation transitions.

An agent keeps its selected action until that action finishes, fails, or
the directive changes, so the policy runs less often than the animation and
agents do not stop and restart on the spot. The interface summarizes the nine needs as five bars, labelled hunger,
thirst, rest, comfort, and social, so that an observer can read an agent's
state at a glance. This grouping affects the display only.

\section{Evaluation}
\label{sec:evaluation}

We organize the evaluation around three questions. \textbf{E1} asks whether
the system correctly identifies instructions that the current scene can
execute. \textbf{E2} asks whether the graph policy selects the intended agents
and responses. \textbf{E3} measures runtime cost and scaling relative to
handwritten rules and per-tick LLM control.

Each question is reported in three parts: how the measurement
was made, what the result shows, and what it implies for the design.

\subsection{Evaluation Setup and Comparison Methods}
\label{sec:evaluation-setup}

\paragraph{Participant reference.}
Fifteen volunteers (11 male and 4 female; age 18--50+) completed an anonymous,
approximately 15-minute web form. A $5\times4$ pool covered five instruction
categories across four venues; each participant received a stratified random
sample of 10 scenarios. Q1 assessed scene feasibility, Q2 identified the
affected agents, and Q3 selected the preferred response from four executable,
scene-specific candidate actions. Answer order was randomized. Scenarios
overlapped across participants, so confidence intervals use a cluster bootstrap
over scenarios.

\begin{table*}[t]
  \centering
  \caption{Policy evaluation. Panel~A compares alternative approaches. Q2
  Jaccard measures agreement on the affected scope; Q3 measures agreement on
  the preferred candidate action. The per-tick LLM is a comparison method
  rather than a local decision maker. Panel~B reports controlled
  ablations with identical data, candidates, masks, labels, and
  optimization. Learned results are means over three seeds. Best value in bold, second best underlined.}
  \label{tab:e2-analysis}
  \small

  \textbf{Panel A: Comparison with alternative approaches}

  \setlength{\tabcolsep}{5.0pt}
  \begin{tabular}{llrrrr}
    \toprule
    Approach & Runtime language use & Q2 Jaccard $\uparrow$ & Q3 $\uparrow$
      & Acceptable $\uparrow$ & Prohibited $\downarrow$ \\
    \midrule
    CrowdDirector & per new instruction & \underline{0.595} & \underline{0.611}
      & \underline{0.479} & \textbf{0.211} \\
    Handwritten rules & none & 0.300 & 0.233 & 0.312 & 0.368 \\
    Per-tick LLM control & every tick & \textbf{0.683} & \textbf{0.667}
      & \textbf{0.566} & \underline{0.217} \\
    \bottomrule
  \end{tabular}

  \textbf{Panel B: Controlled component ablations}

  \setlength{\tabcolsep}{5.5pt}
  \begin{tabular}{lrrrrrr}
    \toprule
    Variant & Q2 Jaccard $\uparrow$ & Q3 $\uparrow$ & Acceptable $\uparrow$
      & Prohibited $\downarrow$ & Aff. recall $\uparrow$ & Aff. F1 $\uparrow$ \\
    \midrule
    CrowdDirector & \textbf{0.595} & \textbf{0.611} & \underline{0.479}
      & \underline{0.211} & \textbf{0.844} & \textbf{0.760} \\
    \quad w/o directive-target encoding
      & \textbf{0.595} & 0.278 & 0.367 & \textbf{0.172} & 0.799 & 0.729 \\
    \quad w/o message passing
      & \underline{0.578} & \underline{0.400} & \textbf{0.482} & 0.344
      & \underline{0.803} & \underline{0.741} \\
    \bottomrule
  \end{tabular}
\end{table*}

\paragraph{Behavioural benchmark.}
We use a frozen benchmark of 100 scenarios, each run under 10
initializations. It is generated once and reused unchanged, so every method is
measured on identical situations. Each scenario states in advance which agents should
respond, and which of the actions open to them the instruction permits. An
agent that takes a permitted action gives an \emph{acceptable} response, and
one that takes an action the user ruled out gives a \emph{prohibited} response,
even though both are physically executable in the scene. The agents that
actually changed behaviour are compared against the intended set by precision,
recall, and F1, so disturbing a bystander and missing an intended agent are
counted separately. Learned variants are averaged over three seeds.

We compare CrowdDirector, which interprets an instruction once and then
decides locally, with two alternatives. A handwritten rule set never uses
language at runtime, and a temperature-zero LLM decides every update from one
batched call covering the whole crowd. E2 uses the matched subset in which all three faced the
same situation, 760 of the 840 cached calls. Fifteen LLM outputs in that subset
could not be parsed, leaving 745 valid rows; where the LLM validly assigned no
agent, default behaviour is retained.

\subsection{E1: Can the Instruction Be Executed?}
\label{sec:e1}

\paragraph{Method.}
For each instruction--scene pair we collect two independent
feasibility judgements and compare both against people. The LLM judges first:
while interpreting the instruction, it also states whether it believes the
scene can execute it. The scene validator then checks the same instruction
against the live scene and returns its five-state verdict, which we reduce to a
binary decision by treating executable and partially executable as supported,
and unsupported, ambiguous, and invalid reference as unsupported. The reference
is the majority Q1 judgement of the participants, and agreement is the
proportion of scenarios on which an assessor matches that reference.

\begin{table}[t]
  \centering
  \caption{Feasibility agreement with participant Q1 judgements.
  Best value in bold.}
  \label{tab:e1-feasibility}
  \small
  \setlength{\tabcolsep}{7pt}
  \begin{tabular}{lcc}
    \toprule
    Feasibility assessor & Agreement $\uparrow$ & 95\% CI \\
    \midrule
    Scene validator & \textbf{0.83} & $[0.61,\,1.00]$ \\
    LLM self-assessment & 0.47 & $[0.21,\,0.72]$ \\
    \bottomrule
  \end{tabular}
\end{table}

\paragraph{Results.}
Table~\ref{tab:e1-feasibility} reports the agreement. The scene validator agrees more closely with participant judgements than
the LLM's self-assessment. The intervals overlap, so the margin alone is
not conclusive, but the direction is consistent: where the two disagree, the
validator is usually the one people side with. The LLM's errors are not random.
It tends to accept instructions that read sensibly but name something the scene
does not contain, which is exactly the case its own confidence cannot detect.

\paragraph{Why separate interpretation from validation.}
These results support separating language interpretation from scene
validation. An LLM can infer the likely meaning of an instruction, but its own
confidence does not establish that the referenced object, affordance, or group
exists, because that is a fact about the scene rather than about the sentence.
Checking the interpretation against the explicit contents of the scene also
produces a clear failure state before an invalid instruction reaches the graph
policy, which is what allows the system to tell the user that part of a request
cannot be carried out rather than silently doing something else.

\begin{table*}[t]
  \centering
  \caption{Runtime and scaling. (a)~separates the sparse cost of
  authoring an instruction from the recurrent cost of control, instantiated for
  a session of $N=5$ instructions and $T=600$ updates. (b)~reports CPU scaling
  at approximately constant agent density. (c)~reports the fraction of
  whole-crowd LLM calls that assigned every agent before reaching the
  output-token limit, over 15 calls per crowd size. Measured on an Intel
  i9-13900H (14 physical cores), CPU only. Best value in bold.}
  \label{tab:e3-scale}
  \small

  \setlength{\tabcolsep}{4.5pt}
  \begin{tabular}{lrrrrrr}
    \toprule
    \multicolumn{7}{l}{\textbf{(a)\;\;Authoring and recurrent control}} \\
    \midrule
    Approach & Calls/instr. & Calls/update & Calls/session
      & Latency/update $\downarrow$ & Cost/update & Cost/session $\downarrow$ \\
    \midrule
    CrowdDirector & 1 & \textbf{0} & $1+N=\textbf{6}$
      & \textbf{20.5}\,ms & \textbf{\$0} & \textbf{\$0.03} \\
    Whole-crowd LLM & 1 & 1 & $1+N+T=606$
      & 3{,}880\,ms & \$0.0038 & \$2.30 \\
    \bottomrule
  \end{tabular}

  \setlength{\tabcolsep}{4.8pt}
  \begin{tabular}{lrrrrrrrr}
    \toprule
    \multicolumn{9}{l}{\textbf{(b)\;\;CrowdDirector computational scaling}} \\
    \midrule
    Crowd size & 20 & 50 & 100 & 200 & 500 & 1{,}000 & 2{,}000 & 5{,}000 \\
    \midrule
    Latency/update (ms) & 20.5 & 36.4 & 64.7 & 113.1
      & 281.2 & 566.7 & 1{,}130 & 3{,}230 \\
    \bottomrule
  \end{tabular}

  \setlength{\tabcolsep}{8.5pt}
  \begin{tabular}{lrrrrr}
    \toprule
    \multicolumn{6}{l}{\textbf{(c)\;\;Whole-crowd LLM assignment completeness}} \\
    \midrule
    Crowd size & 20 & 30 & 40 & 60 & 80 \\
    \midrule
    Complete assignment rate $\uparrow$ & 1.00 & 0.33 & 0.00 & 0.00 & 0.00 \\
    \bottomrule
  \end{tabular}
\end{table*}

\subsection{E2: Do Agents Respond to the Instruction?}
\label{sec:e2}

\paragraph{Method.}
E2 asks two separate questions about a response: whether the right
agents reacted, and whether they did the right thing. Q2 scores the affected
scope as the Jaccard overlap between the agents an approach moves and the
agents participants identified, so both missing an affected agent and moving an
unaffected one are penalized. Q3 scores the chosen response as the proportion
of cases matching the participants' preferred candidate. The benchmark
additionally reports the acceptable and prohibited rates defined above, and
precision, recall, and F1 over the affected agents. Panel~A compares the three
approaches on the matched subset. Panel~B removes one component of
CrowdDirector at a time, holding the data, candidates, masks, labels, and
optimization identical, so a change can be attributed to that component.

\paragraph{Results: comparison with rules and per-tick LLM.}
Panel~A of Table~\ref{tab:e2-analysis} compares the three approaches.
The ordering is the same on every behavioural measure.
Per-tick LLM control is strongest, CrowdDirector is close behind, and the
handwritten rules are far below both. CrowdDirector roughly doubles the rule
set on both scope and response agreement, and it recovers most of the gap
between the rules and an LLM that reasons at every update.
On prohibited responses the two are effectively level, so the
remaining gap is one of coverage rather than safety: CrowdDirector sometimes
fails to find the intended response, but it is no more likely to choose one the
instruction ruled out.
These results match the intended operating point:
CrowdDirector does not reproduce the full flexibility of recurrent LLM
reasoning, but transfers substantially more instruction sensitivity into
local control than a fixed rule set.

\paragraph{Why store an instruction.}
This comparison is the case for storing the instruction rather than
regenerating it. The handwritten rules show what is lost without language at
runtime: they are efficient, but they only cover responses someone anticipated,
and varied instructions fall outside them. Per-tick LLM control shows what
continuous language buys, and it is real but modest relative to its cost, since
it repeats the same inference even when the user's intent has not changed.
Keeping the validated instruction as state captures most of that benefit once,
and leaves the repeated work to a component that can run locally.

\paragraph{Results: component ablations.}
Panel~B isolates two design choices. Removing the directive-target
marker leaves scope agreement untouched but roughly halves response agreement.
The two are therefore decided separately: the marker tells an already-affected
agent which destination the directive assigned, and contributes nothing to
deciding whether that agent is affected at all. Removing message passing
behaves differently. Response agreement falls less sharply, but affected-agent
recall and F1 both drop and the prohibited rate rises by more than half, which
is the clearest single degradation anywhere in the table.

\paragraph{What the graph policy contributes.}
The two components address complementary failures. A policy cannot follow
an assigned destination unless that assignment is visible in its candidate
representation, which is what the directive-target marker supplies and why its
removal affects only the response. Relational context does something the marker
cannot: it connects an agent's decision to the surrounding event, object, and
group structure, and without it the policy more often selects a response the
instruction ruled out. One ablation is worth noting against us. Removing the
directive-target marker lowers the prohibited rate slightly, because a policy
that less often pursues the assigned destination also less often reaches the
situation where a prohibited alternative is attractive. The marker improves
instruction following, and that is what it is for, but it does not by itself
make responses safer.

\subsection{E3: Runtime Cost and Scaling}
\label{sec:e3}

\paragraph{Method.}
E3 tests the central runtime premise of CrowdDirector: language is used when
the scene or director's intent changes, while repeated crowd-control updates
are performed locally. Let $N$ denote the number of user instructions and $T$
the number of decision updates in a session. Including one initial
scene-generation call, the API-call counts are
\begin{equation}
A_{\mathrm{CD}}=1+N, \qquad A_{\mathrm{LLM}}=1+N+T.
\label{eq:runtime-calls}
\end{equation}
Both configurations retain language-based authoring. CrowdDirector removes the
recurrent $T$ calls from the crowd-control loop. The difference is small
symbolically and large in practice, because $T$ is the length of the session
while $N$ is only the number of times the user speaks: for five instructions
over 600 updates, six calls against 606.
We measure latency per complete decision update on identical 20-agent
worlds, and price each call at the provider's published rate. Panel~B then
raises the crowd size at approximately constant density to separate the cost of
the graph policy from the cost of language. Panel~C reports how often the
whole-crowd LLM returned a parseable assignment for every agent, over 15 calls
at each crowd size.

\paragraph{Results.}
Table~\ref{tab:e3-scale} reports runtime and scaling. Removing language from the control loop reduces per-update latency by
$189\times$, and over a session of five instructions and 600 updates it reduces
estimated cost by nearly two orders of magnitude. Both configurations pay the
same price to author an instruction; the difference is entirely in what
recurs.

Local runtime grows close to linearly with crowd size and adds no further
API calls, staying interactive well past the range the policy was trained on.
The whole-crowd LLM interface fails for a different reason: because it
enumerates one assignment per agent, its output grows with the crowd and it
stops returning complete responses shortly above the smallest size tested. Its
limit is the response format, not the reasoning. Multiple calls or a grouped
representation would move it, and CrowdDirector takes the compressed route by
naming the event, scope, and target once and expanding it locally.

The policy was trained on crowds of 8--40 agents. The large-crowd experiment
therefore demonstrates computational scaling, not behavioural generalization.
It shows that the crowd graph and graph policy can be evaluated beyond the training
range without language cost; it does not establish that response quality is
unchanged for thousands of agents.

\paragraph{Connecting decisions to Unity execution.}
These savings depend on the boundary the method draws. The Unity
implementation is not a separate decision maker but the execution side of the same
pipeline: the graph policy selects an action and target, and the engine plans
movement, carries the action out, and reports completion or failure. Keeping
that boundary explicit is what makes the recurrent loop cheap, since the
learned component never has to produce physically valid trajectories, and
established navigation and animation systems continue to do the work they
already do well.

\section{Conclusion and Future Work}
\label{sec:conclusion}

We presented CrowdDirector, a framework for directing a running crowd
simulation with natural language. An LLM converts a new instruction into a
structured instruction, the scene validator checks it against the current
scene, the accepted directive remains active, and the graph policy selects a
candidate action for each agent. Unity then executes those choices with its
navigation and interaction systems. The evaluation shows that explicit scene
validation improves agreement with participant judgements and that the graph
policy follows instructions more closely than handwritten rules without
requiring language inference at every update. Per-tick LLM control achieves
higher response agreement in our benchmark, but with substantially higher
recurrent latency and API cost. CrowdDirector therefore provides a practical
middle ground between flexible instruction following and efficient runtime
crowd control.

However, we identify several limitations of the present study as follows: (1)~the current implementation supports only the actions and interactions defined in its operation library and has been evaluated in 2D environments; (2)~the participant study involved 15 volunteers and a limited set of instruction scenarios; and (3)~the policy was trained with crowds of 8--40 agents, so the larger-crowd experiment evaluates computational scaling rather than behavioral generalization beyond the training range. Future work will extend CrowdDirector
to richer environments and interactions, evaluate a broader range of
participants and instructions, and improve large-crowd efficiency.

\begin{acks}
This work was conducted with the financial support of the Research Ireland
Centre for Research Training in Digitally-Enhanced Reality (d-real) under Grant
No.~18/CRT/6224. For the purpose of Open Access, the author has applied a CC BY
public copyright licence to any Author Accepted Manuscript version arising from
this submission.
\end{acks}

\bibliographystyle{ACM-Reference-Format}
\bibliography{sample-base}

\end{document}